\documentstyle[aps,12pt,epsf]{revtex}
\def\la{\mathrel{\mathpalette\fun <}}
\def\ga{\mathrel{\mathpalette\fun >}}
\def\fun#1#2{\lower3.6pt\vbox{\baselineskip0pt\lineskip.9pt
  \ialign{$\mathsurround=0pt#1\hfil##\hfil$\crcr#2\crcr\sim\crcr}}}
\begin{document}

\draft
\title{Leptonic Domains in the Early Universe and Their Implications}

\author{Xiangdong~Shi and George~M.~Fuller}
\address{Department of Physics, University of California,
San Diego, La Jolla, California 92093-0319}

\date{January 16, 1998}

\maketitle

\begin{abstract}
We extend a treatment of the causal structure of space-time to
active-sterile neutrino transformation-based schemes for lepton
number generation in the early universe. We find that these
causality considerations necessarily lead to the creation of
spatial domains of lepton number with opposite signs.  Lepton
number gradients at the domain boundaries can open a new channel
for MSW resonant production of sterile neutrinos.  The enhanced
sterile neutrino production via this new channel allows considerable
tightening of Big Bang Nucleosynthesis constraints on active-sterile
neutrino mixing, including the proposed $\nu_\mu\rightarrow\nu_s$ 
solution for the Super Kamiokande atmospheric $\nu_\mu$ deficit, 
and the four-neutrino schemes proposed to simultaneously fit current
neutrino experimental results.
\end{abstract}
\bigskip


\newpage
It is well known that resonant MSW (Mikheyev-Smirnov-Wolfenstein)
\cite{MSW} transitions between active neutrinos and sterile
neutrinos in the early universe could generate large lepton number
asymmetries in the neutrino sector \cite{Foot1,Shi2,Foot2,Shi3}.
Here the lepton number for an active neutrino species $\nu_\alpha$
is defined to be $L_{\nu_\alpha}\equiv (n_{\nu_\alpha}
-n_{\bar\nu_\alpha})/n_\gamma$, the net asymmetry in $\nu_\alpha$
over $\bar\nu_\alpha$ number density normalized by the photon
number density $n_\gamma$.

Not well noted is a crucial feature involved
in the lepton number generation process\cite{Shi2,Shi3}:
that the lepton number asymmetry is first damped to
essentially zero by the active-sterile neutrino mixing, 
then oscillates chaotically with an progressively larger
amplitude as the mixing goes through resonances, until 
the asymmetry converges to a growing asymptotic value 
that is either positive or negative. As a result of this
feature, the sign of the lepton number asymmetry is
independent of the initial conditions which obtain before 
the instability begins, and is exponentially sensitive 
to the parameters involved during the chaotic oscillitory
phase.  In turn, the lepton number generated in this process
may not have a uniform sign in different causal domains.  
Obviously, an upper bound to the size of these domains
is the particle horizon $H^{-1}\approx (90/8\pi^3)^{1/2}
g^{-1/2}m_{\rm pl}/T^2$ where $g$ is the statistical weight
in relativistic particles, $T$ is the temperature of the
universe, and $m_{\rm pl}\approx 1.22\times 10^{28}$ eV.

The typical size of these domains is at least as large as the
diffusion length of neutrinos at the time of the lepton number
generation.  These leptonic domains can persist as long as the
resonant neutrino transition is capable of efficient lepton number
generation. This is because any reduction of lepton number at
domain boundaries due to mixing will be quickly reversed by the
generation process so that a boundary region is incorporated into
one domain or the other.

The existence of leptonic domains in the early universe provide
a new channel of producing sterile neutrinos, via the resonant 
MSW conversion of active neutrinos to sterile neutrinos at domain
boundaries, where lepton number gradients exist.
If during the Big Bang Nucleosynthesis (BBN) epoch the population
of sterile neutrinos from this channel becomes comparable to that
of an equilibrated active neutrino flavor, the extra neutrino 
degrees-of-freedom (an increase in $g$) can increase
the primordial helium abundance significantly \cite{Schramm}.
Therefore, such a production mechanism can be constrained by
the observationally-inferred primordial helium abundance.

The details of the active-sterile neutrino transformation process
in the early universe and the associated generation of lepton
number asymmetries can be found in a number of previous works
\cite{Shi2,Foot2,Shi3}. Here we only briefly summarize. 
In all calculations, we employ the natural units $\hbar=c=k_{\rm B}=1$. 
A two-family system $\nu_\alpha\leftrightarrow\nu_s$ ($\alpha = e$,
$\mu$ or $\tau$, and $\nu_s$ is the sterile neutrino) has a 2$\times$2
evolution Hamiltonian ${\cal H}$
with ${\cal H}_{\alpha\alpha}=V_z$, 
     ${\cal H}_{\alpha s}={\cal H}_{s \alpha}^\dag=V_x+iV_y$,
     ${\cal H}_{ss}=0$.
The effective vector potential {\bf V} during
the BBN epoch (below the QCD phase transition
temperature $T\la 100$ MeV) is
\begin{equation}
V_x=-{\delta m^2\over 2E}\sin 2\theta,\quad
V_y=0,\quad
V_z=-{\delta m^2\over 2E}\cos 2\theta+V_\alpha^L+V_\alpha^T,
\label{potential}
\end{equation}
where $\delta m^2\approx m_{\nu_s}^2-m_{\nu_\alpha}^2$,
$\theta$ is the vacuum mixing angle, and $E$ is neutrino energy.
The matter-antimatter asymmetry contribution to the effective
potential is\cite{Mattereffect}
\begin{equation}
V_\alpha^L\approx \pm 0.35 G_FT^3\Bigl[L_0+2L_{\nu_\alpha}+
                  \sum_{\beta\ne\alpha}L_{\nu_\beta}\Bigr]
\label{VL}
\end{equation}
with the \lq\lq $+$\rq\rq\  sign for $\nu_\alpha$
and the \lq\lq $-$\rq\rq\  sign for $\bar\nu_\alpha$.
The quantity $L_0$ represents the contribution from the baryonic 
asymmetry and electron-positron asymmetry, i.e., $L_0\sim 10^{-10}$.
The quantity $L_{\nu_\beta}$ is the asymmetry in other active 
neutrino species $\nu_\beta$. For simplicity, and with no loss
of generality, we will assume $L_{\nu_\beta}=0$ unless explicitly
stated otherwise. The contribution to {\bf V} from a thermal
neutrino background is \cite{Mattereffect}
\begin{equation}
V_\alpha^T \approx \left\{\begin{array}{ll}
 -80G_F^2ET^4 & \mbox{for $\alpha=e$;}\\
 -22G_F^2ET^4 & \mbox{for $\alpha=\mu,\,\tau$.}\end{array}
\right.
\end{equation}
$V_\alpha^T$ is the same for both $\nu_\alpha$ and $\bar\nu_\alpha$.

Resonances occur when the diagonal elements of the Hamiltonian
${\cal H}$ are equal, i.e., $V_z=0$. This is when $\nu_\alpha$ 
and $\nu_s$ become degenerate in effective mass and maximally mixed
in medium. Very simplistically, when $V_z$ evolves through zero,
$\nu_\alpha$ and $\nu_s$ swap their flavors if the resonance is
adiabatic ($\vert{\rm d}V_z/{\rm d}t\vert\ll V_x^2$), and remain
essentially unaltered if the resonance is non-adiabatic 
($\vert{\rm d}V_z/{\rm d}t\vert\gg V_x^2$).  Since the resonance
condition is energy dependent, only neutrinos with $E_{\rm res}$
are are resonant at any given temperature.

When $V_\alpha^L=0$ (or negligibly small compared to 
$\vert\delta m^2\vert/2E$), resonances can occur
for the $\nu_\alpha\leftrightarrow\nu_s$ system only
if $\delta m^2<0$ (i.e., $m_{\nu_\alpha}>m_{\nu_s}$).
For $\delta m^2>0$, and at high enough temperatures for $\delta m^2<0$,
it has been shown that active-sterile neutrino transformation can
damp $L_0+2L_{\nu_\alpha}$ to zero very efficiently. The amplification
of $L_0+2L_{\nu_\alpha}$ starts for $\delta m^2<0$ only as the 
temperature falls below the critical temperature
\begin{equation}
T_{\rm c}\approx 22\left\vert\delta m^2/1\,{\rm eV}^2\right\vert^{1/6}\,{\rm MeV}.
\end{equation}
This is also the temperature where a significant number of $\nu_\alpha$
go through the resonance. The instability of $L_0+2L_{\nu_\alpha}$ at
$T_{\rm c}$ results from the non-linear characteristics of the MSW
resonances, and the feedback effect they have on the lepton number
asymmetry\cite{Shi2}.

Due to the same non-linearity, $L_0+2L_{\nu_\alpha}$
(or $L_{\nu_\alpha}$ once $L_{\nu_\alpha}\gg L_0$) 
oscillates chaotically near $T_c$, with an exponentially
increasing amplitude. 
This chaotic feature was first pointed out by Shi in 
numerical studies of monochromic neutrinos undergoing
MSW resonance\cite{Shi2}, and was later found to apply
to systems with a distribution of energies as well \cite{Shi3}. 
Eventually, at a temperature slightly below $T_c$, the
oscillatory behavor abates, and $L_{\nu_\alpha}$ settles
into one of two fixed points (two $T^{-4}$ power-laws)
\cite{Foot1,Shi2,Foot2,Shi3}:
\begin{equation}
L^{(\pm)}_{\nu_\alpha}\sim 
\pm\left\vert{\delta m^2\over 10\,{\rm eV}^2}\right\vert
\left({T\over 1\,{\rm MeV}}\right)^{-4}.
\label{powerlaw}
\end{equation}
The sign of the emergent $L_{\nu_\alpha}$, however,
is independent of the initial $L_0$ and $L_{\nu_\alpha}$
above $T_c$, and is exponentially sensitive to the
evolution of the potential {\bf V} during the oscillatory
phase. The sign is therefore chaotic and uncorrelated
across causal domains.

In the power-law regime, the growth of $L_{\nu_\alpha}$
is driven by the resonant conversion of $\nu_\alpha$
to $\nu_s$ (if $L_{\nu_\alpha}<0$) or $\bar\nu_\alpha$
to $\bar\nu_s$ (if $L_{\nu_\alpha}>0$) in the sector of
the neutrino energy spectrum with ${E_{\rm res}/T}
\approx 0.06\vert\delta m^2/1\,{\rm eV}^2\vert
\,\vert L_{\nu_\alpha}\vert^{-1}(T/1\,{\rm MeV})^{-4}$,
which is in general $\ll 1$\cite{Foot2,Shi3}. In this
regime, $V^T_\alpha$ quickly becomes negligible
because it scales as $T^5$.  The power-law solution,
Eq.~(\ref{powerlaw}), for $\vert L_{\nu_\alpha}\vert$
is very stable, in that a $\vert L_{\nu_\alpha}\vert$
significantly deviating from this solution will be 
quickly damped or amplified until it converges to 
Eq.~(\ref{powerlaw}).
(A larger $\vert L_{\nu_\alpha}\vert$ implies a small $E_{\rm res}/T$,
and therefore a less efficient generation of $\vert L_{\nu_\alpha}\vert$;
a smaller $\vert L_{\nu_\alpha}\vert$ implies a larger $E_{\rm res}/T$
and a more efficient generation of $\vert L_{\nu_\alpha}\vert$.)
The rate of this convergence is ${\rm d}\ln\vert L_{\nu_\alpha}\vert/{\rm d}
\ln T\ga 4$ in the power-law regime. 

As $\vert L_{\nu_\alpha}\vert$ increases, $E_{\rm res}/T$ will
slowly increase.  This is because keeping up with the approximate $T^{-4}$ 
growth requires progressively more $\nu_\alpha$ or $\bar\nu_\alpha$
to be resonantly converted into sterile neutrinos. Eventually, as 
$E_{\rm res}/T$ sweeps through most of the $\nu_\alpha$ energy
spectrum, the growth of $\vert L_{\nu_\alpha}\vert$ tapers off near 
its physical limit. This limit is $\vert L_{\nu_\alpha}\vert=3/8$,
when the entire $\nu_\alpha$ or $\bar\nu_\alpha$ population has 
been converted into sterile neutrinos. In turn, the lepton number
generation process ceases, and any inhomogeneity of lepton number
arising from the process begins to be smoothed out by neutrino diffusion.

Because of the chaotic behavior of the sign of $L_{\nu_\alpha}$,
domains of lepton number with opposite signs are expected to 
form at the epoch when $T\approx T_c$. At the domain boundaries,
mixing between different domains tends to reduce the asymmetries
and increases the thickness of the boundary regions. However, the
resonant neutrino mixing tends to maintain the solution in
Eq.~(\ref{powerlaw}) and so narrows the boundary region.
The thickness of the boundaries, very crudely, is thus
the diffusion length of active neutrinos within the time
in which $\vert L_{\nu_\alpha}\vert$ grows by an $e$-folding,
$H^{-1}/4$. Taking the $\nu_\alpha$ collision rate
$\Gamma_{\nu_\alpha}\sim G_F^2T^5$, we obtain
a boundary thickness relative to the horizon scale $ct$
\begin{equation}
\delta_d\sim {c\over \Gamma_{\nu_\alpha}}
\left({\Gamma_{\nu_\alpha}\over 4H}\right)^{1/2}\sim ct
\left({T\over 1\,{\rm MeV}}\right)^{-1.5},
\end{equation}
for $T\ga 1$ MeV when active neutrinos are still diffusive. 
At temperatures $T\la 1$ MeV, neutrinos decouple from the plasma
and free-stream at the speed of light. In this case $\delta_d\sim ct$.

The existance of lepton domains results in a gradient of
$L_{\nu_\alpha}$, and therefore a gradient in $V_z$. More
importantly, the varying $V_z$ at the domain boundaries
satisfies the resonant condition for most $\nu_\alpha$ 
crossing the boundaries.  This is because [as Eq.~(\ref{powerlaw})
shows] the power-law solutions within
each domain result from resonant transitions of $\nu_\alpha$ or 
$\bar\nu_\alpha$ with $E_{\rm res}\ll T$\cite{Foot2,Shi3}.
As a result, at domain boundaries $E_{\rm res}$ becomes larger.
This is because $E_{\rm res}\propto \vert L_{\nu_\alpha}\vert^{-1}$.
Most of $\nu_\alpha$ and $\bar\nu_\alpha$ therefore
undergo resonant transitions to sterile neutrinos
at domain boundaries. 

This new channel of sterile neutrino production can 
populate a significant sterile neutrino sea with a 
number density comparable to that of an equilibrated
active neutrino flavor if (1) the resonant conversion
at the boundary region is adiabatic; and (2) this new
production channel for sterile neutrinos does not provide
negative feedback to the lepton number asymmetry generation
process, and so does not compromise the domain structure of
lepton number.  Here we demonstrate that both requirements
can be satisfied.

The adiabaticity condition at the resonances is
\begin{equation}
V_x^2>\left\vert {{\rm d}V_z\over {\rm d}t}\right\vert=
\left\vert c{\partial V_z\over \partial r}-HT{\partial 
V_z\over \partial T}\right\vert,
\label{adiabaticity}
\end{equation}
evaluated at $V_z=0$.
The second term on the r.h.s. is of order 
$HV_\alpha^L$. This term is always small 
compared to the first term as long as the
leptonic domains exist.  Since the spatial
gradient is expected to be smooth across the
boundary (whose thickness is determined by the
diffusion/streaming process), we can employ the
average $\vert\partial V_z/\partial r\vert$
across the domain boundaries, $\sim 0.7G_FT^3
(L^{(+)}_{\nu_\alpha}-L^{(-)}_{\nu_\alpha})/\delta_d$
[where $L^{(+)}_{\nu_\alpha}\sim -L^{(-)}_{\nu_\alpha}$
satisfying Eq.~(\ref{powerlaw})].  In turn,
the adiabaticity condition Eq.~(\ref{adiabaticity}) becomes
\begin{equation}
\left\vert\delta m^2\right\vert^2\sin^22\theta > 10H
\left({E\over T}\right)^2G_FT^5\left({T\over 1\,{\rm MeV}}\right)^{1.5}\,
{\rm min}\left[\left\vert{\delta m^2\over 10\,{\rm eV}^2}\right\vert
\left({T\over 1\,{\rm MeV}}\right)^{-4},{3\over 8}\right].
\label{master}
\end{equation}

The production of sterile neutrinos via adiabatic
conversion of $\nu_\alpha$ at domain boundaries
will not have a negative impact on the $L_{\nu_\alpha}$
generation process and the domain structure of lepton
number.  Consider a domain boundary with a lepton 
asymmetry $L^{(+)}_{\nu_\alpha}$ on one side, and
$L^{(-)}_{\nu_\alpha}$ on the other. The $L^{(+)}_{\nu_\alpha}$ 
side has fewer $\bar\nu_\alpha$ than $\nu_\alpha$, and
some $\bar\nu_s$ from resonant $\bar\nu_\alpha\rightarrow\bar\nu_s$
conversions.  The $L^{(-)}_{\nu_\alpha}$ side has the opposite,
with more $\bar\nu_\alpha$ than $\nu_\alpha$, and
some $\nu_s$ from resonant $\nu_\alpha\rightarrow\nu_s$
conversions.  When neutrinos cross the boundary from the
$L^{(+)}_{\nu_\alpha}$ side to the $L^{(-)}_{\nu_\alpha}$
side, the resultant production of sterile neutrinos due to
$\nu_\alpha\rightarrow\nu_s$ and 
$\bar\nu_\alpha\rightarrow\bar\nu_s$ has no bearing on
the asymmetry of $\nu_\alpha$ on the $L^{(-)}_{\nu_\alpha}$ side.
The existence of sterile neutrinos do not hinder
the $L_{\nu_\alpha}$ generation process until the
sterile neutrino population is comparable in numbers
to the $\nu_\alpha$ population\cite{Shi2}.
The resonant conversion of $\bar\nu_s\rightarrow\bar\nu_\alpha$
due to the crossing, on the other hand, produces more
$\bar\nu_\alpha$ in the $L^{(-)}_{\nu_\alpha}$ domain
and only reinforces the domain structure.

Therefore, once the adiabaticity condition Eq.~(\ref{master})
is met, this new channel of sterile neutrino production
may be potent enough to bring the sterile neutrinos into
equilibrium with active neutrinos.  (In fact, this condition
is conservative because neutrinos may cross multiple domain 
boundaries within a Hubble time.)  On the other hand,
the observationally-inferred primordial $^4$He abundance
\cite{Olive,Thuan} and deuterium abundance \cite{Tytler}
constrain the total number of neutrino flavors in equilibium
$N_\nu$ to be $\la 3.3$ \cite{Shi3,Schramm,Turner} ($N_\nu$
represents relativistic degrees of freedom in neutral
fermions). This constraint implies that the new sterile
neutrino production channel
cannot be efficient before the decoupling of the
$\nu_\alpha\bar\nu_\alpha$ pair production process.
For $\alpha=\mu$ and $\tau$, this 
decoupling temperature is $\sim 5$ MeV. A constraint on 
the two-family $\nu_{\mu,\tau}\leftrightarrow\nu_s$ mixing
can therefore be obtained by requiring that the adiabaticity 
condition Eq.~(\ref{master}) is not satisfied at
$T\sim {\rm max}(5,\vert\delta m^2/4\,{\rm eV}^2\vert^{1/4})$ MeV
(the latter term in the bracket is the temperature at which
the growth of $L_{\nu_\alpha}$ stops):
\begin{equation}
\begin{array}{ll}
\left\vert\delta m^2\right\vert\sin^22\theta < 7\times 10^{-5}\,{\rm eV}^2 &
\mbox{for $\left\vert\delta m^2\right\vert \la 2.5\times 10^3\,{\rm eV}^2$;}\\
\sin^22\theta < 3\times 10^{-8} &
\mbox{for $\left\vert\delta m^2\right\vert \ga 2.5\times 10^3\,{\rm eV}^2$.}
\end{array}
\end{equation}

For $\alpha=e$, the constraint from BBN is more severe. 
The sterile neutrino production cannot be efficient not
only above the $\nu_e\bar\nu_e$ pair production decoupling
temperature $T\sim 3$ MeV,
but also at the weak freeze-out temperature of $T\sim 1$ MeV. 
At this temperature, a significant $\nu_e\rightarrow\nu_s$
and $\bar\nu_e\rightarrow\bar\nu_s$ transition would
cause a deficit in the $\nu_e\bar\nu_e$ number density,
which cannot be replenished by pair production.  A significant
deficit in the $\nu_e\bar\nu_e$ number density causes the
neutron-to-proton ratio to freeze out too early and results
in a primordial $^4$He abundance that is too large. (For
example, a 10\% deficit in the $\nu_e\bar\nu_e$ number density
has roughly the same effect on the $^4$He abundance as $N_\nu\approx3.5$.)
Therefore the BBN constraint on the two-family
$\nu_e\leftrightarrow\nu_s$ mixing is obtained at $T\sim 
{\rm max}(1,\vert\delta m^2/4\,{\rm eV}^2\vert^{1/4})$ MeV:
\begin{equation}
\begin{array}{ll}
\left\vert\delta m^2\right\vert\sin^22\theta < 5\times 10^{-8}\,{\rm eV}^2 &
\mbox{for $\left\vert\delta m^2\right\vert \la 4\,{\rm eV}^2$;}\\
\sin^22\theta < 10^{-8} &
\mbox{for $\left\vert\delta m^2\right\vert \ga 4\,{\rm eV}^2$.}
\end{array}
\end{equation}

These bounds are summarized in Figure 1. They apply in addition
to the previous bounds based on a universe with homogeneous lepton
numbers, and together they offer much tighter constraints on
the two-family active-sterile neutrino mixing.

Intriguing results can also be obtained if there is
active-sterile neutrino mixing involving three or
more families. One example is the proposal that a
$\nu_\mu\leftrightarrow\nu_s$ and
$\nu_\tau\leftrightarrow\nu_{s^\prime}$ (in principle 
$\nu_{s^\prime}$ and $\nu_s$ can be the same flavor)
mixing might be able to simultaneously explain
the Super Kamiokande atmospheric neutrino data and 
satisfy the BBN bound\cite{Foot2,Shi3}. A stand-alone
$\nu_\mu\leftrightarrow\nu_s$ oscillation
solution to the Super Kamiokande data would
violate the BBN bound by bringing $\nu_s$ 
into equilibrium during BBN\cite{Enqvist,Shi1}.
The double active-sterile neutrino oscillation
proposal argues that this violation of BBN
bound may be avoided if a resonant
$\nu_\tau\leftrightarrow\nu_{s^\prime}$
transformation in the early universe generates a
$L_{\nu_\tau}$ that hinders the $\nu_s$
production from the $\nu_\mu\leftrightarrow\nu_s$
mixing (by creating a $L_{\nu_\beta}$ term
in Eq.~[\ref{VL}]). This argument is no longer valid
once we consider the existence of the $L_{\nu_\tau}$
domains as a result of the resonant 
$\nu_\mu\leftrightarrow\nu_{s^\prime}$ transformation.
Rather the contrary is true.  The $L_{\nu_\tau}$
domains {\sl facilitate} the production of
$\nu_s$ via resonant $\nu_\mu\rightarrow\nu_s$
transformation at domain boundaries. The adiabaticity
condition, Eq.~(\ref{master}), is modified in this
double mixing situation to be:
\begin{equation}
\left\vert\delta m^2_1\right\vert^2\sin^22\theta_1 >
5H\left({E\over T}\right)^2
G_FT^5\left({T\over 1\,{\rm MeV}}\right)^{1.5}\,
{\rm min}\left[\left\vert{\delta m^2_2\over 10\,{\rm eV}^2}\right\vert
\left({T\over 1\,{\rm MeV}}\right)^{-4},{3\over 8}\right].
\label{modmaster}
\end{equation}
where $\delta m_1^2\approx\vert m_{\nu_\mu}^2-m_{\nu_s}^2\vert$,
$\theta_1$ is the $\nu_\mu\leftrightarrow\nu_s$ vacuum mixing
angle, and $\delta m_2^2\approx m_{\nu_\tau}^2-m_{\nu_{s^\prime}}^2$.
To be consistent with BBN, the adiabaticity condition cannot
be satisfied for the double mixing system from the onset of the
$L_{\nu_\tau}$ generation $T_c\approx 22\vert\delta m_2^2/1
{\rm eV}^2\vert^{1/6}$
MeV to the $\nu_{\mu,\tau}$ decoupling temperature $T\sim 5$
MeV. However, for $\delta m_1^2\sim 10^{-3}$ to $10^{-2}$ eV$^2$ and
$\sin^22\theta_1\sim 1$ (the parameters required to explain the Super
Kamiokande data), the adiabaticity condition is always satisfied in this
double mixing proposal for any reasonable choices of the tau neutrino mass.
Therefore, {\sl BBN unambiguously rules out an active-sterile
neutrino oscillation explanation to the Super Kamiokande data.}

Another interesting situation involving multi-family
active-sterile neutrino mixing arises from neutrino mixing
schemes proposed to explain simultaneously the Los Alamos
Liquid Scintillator Neutrino Detector (LSND) $\bar\nu_e$
signal, Super Kamiokande atmospheric $\nu_\mu$ deficit,
and solar neutrino deficit\cite{Caldwell}. In these models,
$\nu_\mu\leftrightarrow\nu_e$ mixing (with
$m_{\nu_\mu}^2-m_{\nu_e}^2\sim 0.1$ to 10 eV$^2$ and
$\sin^22\theta_{\mu e}\sim 10^{-3}$) is employed to explain
the LSND result and $\nu_e\leftrightarrow\nu_s$ mixing
(with $m_{\nu_s}^2-m_{\nu_e}^2\sim 10^{-5}$ eV$^2$ and
$\sin^22\theta_{es}\sim 10^{-3}$ for the MSW 
solution, and $\vert m_{\nu_s}^2-m_{\nu_e}^2\vert\sim
10^{-10}$ eV$^2$ and $\sin^22\theta_{es}\sim 1$ for
the vacuum solution) is invoked to explain the solar neutrino data.
If there is mixing between $\nu_\mu$ and $\nu_s$ as well,
however, with $\sin^22\theta_{\mu s}\ga 10^{-11}$\cite{Shi2,Foot2},
the $L_{\nu_\mu}$ background and domains generated by
the mixing in the early universe would imply that the
$\nu_e\leftrightarrow\nu_s$ mixing would not populate
enough $\nu_s$ to violate BBN constraints only if
\begin{equation}
\left\vert m_{\nu_s}^2-m_{\nu_e}^2\right\vert^2\sin^22\theta_{es}
<2\times 10^{-14}\left\vert{m_{\nu_\mu}^2-m_{\nu_s}^2\over 1\,{\rm eV}^2}
\right\vert^{1/4}\,{\rm eV}^2\sim 2\times 10^{-14}\,{\rm eV}^2.
\label{es}
\end{equation}
(This is in analogy to the previous example if we take
$\delta m_1^2\approx\vert m_{\nu_s}^2-m_{\nu_e}^2\vert$ and
$\delta m_2^2\approx\vert m_{\nu_\mu}^2-m_{\nu_s}^2\vert$ in
Eq.~[\ref{modmaster}].) This requirement is not satisfied
by the MSW $\nu_e\leftrightarrow\nu_s$ solution to the 
solar neutrino problem. (Note that if the solar neutrino
problem is solved by a vacuum $\nu_e\leftrightarrow\nu_s$ mixing, 
with $\vert m_{\nu_s}^2-m_{\nu_e}^2\vert\sim 10^{-10}$ eV$^2$,
Eq.~(\ref{es}) will be satisfied and the $\nu_e\leftrightarrow\nu_s$
resonant transition will not be adiabatic.)

Therefore, in light of LSND,
Super Kamiokande and solar neutrino experiments,
the neutrino oscillation explanations of the LSND
data and the MSW solution to solar neutrino data 
is inconsistent with BBN unless the 
$\nu_\mu\leftrightarrow\nu_s$ mixing is extremely small,
$\sin^22\theta_{\mu s}\la 10^{-11}$. This results holds despite
the possibility that the $\nu_\mu\leftrightarrow\nu_e$
mixing amplitude and the $\nu_e\leftrightarrow\nu_s$
mixing amplitude could be $\ga 10^8$ times larger. This
restriction severly constrains the $\nu_\tau$-$\nu_\mu$-$\nu_e$-$\nu_s$ 
mixing matrix required to fit the current neutrino experiment results.

In summary, we have discussed the existence of leptonic domains
as an inevitable consequence of resonant active-sterile
neutrino oscillation mechanisms for generation of lepton number.
Resonant MSW conversion due to the lepton number gradients
at domain boundaries therefore provides a new channel for
sterile neutrino production. As a result, the Big Bang
Nucleosynthesis constraint on active-sterile neutrino
mixing becomes much more stringent.  Likewise for the constraint
on multi-family neutrino mixing schemes involving sterile
neutrinos. We have found that the $\nu_\mu\rightarrow\nu_s$
explanation to the Super Kamiokande data is inconsistent
with Big Bang Nucleosynthesis in spite of lepton 
number asymmetries generated by other active-sterile 
neutrino oscillations. We have also found that together the
$\nu_\mu\leftrightarrow\nu_e$ explanation of the LSND
result and the MSW $\nu_e\leftrightarrow\nu_s$ solution
to the solar neutrino problem are incompatible with Big
Bang Nucleosynthesis considerations unless the amplitude
of the mixing between $\nu_\mu$ and $\nu_s$ is $\ga 10^8$
smaller than that between $\nu_\mu$ and $\nu_e$ and that
between $\nu_e$ and $\nu_s$.

X.~S. and G.~M.~F. are supported in part
by NSF grant PHY98-00980 at UCSD.

\newpage
\noindent{\bf Figure Captions:}

\noindent
Figure 1. Parameter spaces to the right of the hatched lines are
excluded by BBN. The solid lines indicate bounds obtained in this
work, and the dashed lines are previous bounds assuming a universe
with a homogeneous lepton number\cite{Shi3,Shi1}.
\bigskip


\begin{references}

\bibitem{MSW} L.~Wolfenstein, {\sl Phys. Rev.} {\bf D17}, 2369 (1978);
		{\sl ibid.} {\bf 20}, 2634 (1979); S.~P.~Mikheyev
		and A.~Yu.~Smirnov, {\sl Sov. Phys. JETP}, {\bf 64}, 4 (1986).

\bibitem{Foot1} R.~Foot, M.~J.~Thomson, and R.~R.~Volkas,
                 {\sl Phys. Rev.} {\bf D 53}, 5349 (1996).

\bibitem{Shi2} X.~Shi, {\sl Phys. Rev.}, {\bf D 54}, 2753 (1996).

\bibitem{Foot2} R.~Foot and R.~R.~Volkas, {\sl Phys. Rev.} {\bf D 55},
                5147 (1997).

\bibitem{Shi3} X.~Shi and G.~M.~Fuller, {\sl Phys. Rev.}, {\bf D}, in press.

\bibitem{Schramm} G.~Steigman, D.~N.~Schramm and J.~E.Gunn, {\sl Phys. Lett}
                  {\bf B 66}, 262 (1977); for a recent update, see
                  C.~Y.~Cardall and G.~M.~Fuller, {\sl Astrophys. J.}
                  {\bf 472}, 435 (1996).

\bibitem{Enqvist} K.~Enqvist, K.~Kainulainen, and M.~Thomson,
                  {\sl Nucl. Phys.} {\bf B 373}, 498 (1992).

\bibitem{Shi1} X.~Shi, D.~N.~Schramm, and B.~D.~Fields, {\sl Phys. Rev.}
               {\bf D 48}, 2568 (1993), and references therein.

\bibitem{McKeller} B.~H.~J.~McKeller and M.~J.~Thomson,
                   {\sl Phys. Rev.} {\bf D 49}, 2710 (1994).

\bibitem{Mattereffect} D.~N\"otzold and G.~Raffelt,
                   {\sl Nucl. Phys.} {\bf B 307}, 924 (1988).

\bibitem{Olive} K.~A.~Olive, E.~Skillman, and G.~Steigman,
		{\sl Astrophys. J.}, {\bf 483}, 788 (1997).

\bibitem{Thuan} Y.~I.~Izotov and T.~X.~Thuan,
		{\sl Astrophys. J.}, {\bf 500}, 188 (1998).

\bibitem{Tytler} S.~Burles and D.~Tytler, to appear in the {\sl
	Proceedings of the Second Oak Ridge Symposium on Atomic \& Nuclear
	Astrophysics}, ed. A. Mezzacappa (Institute of Physics, Bristol),
	and references therein.

\bibitem{Turner} S.~Burles, K.~M.~Nollett, J.~N.~Truran and M.~S.~Turner,
	       submitted to {\sl Phys. Rev. Lett.}

\bibitem{Caldwell} D.~O.~Caldwell, {\sl Int. J. Mod. Phys.}
		{\bf A13} 4409 (1998), and references therein.
\end{references}
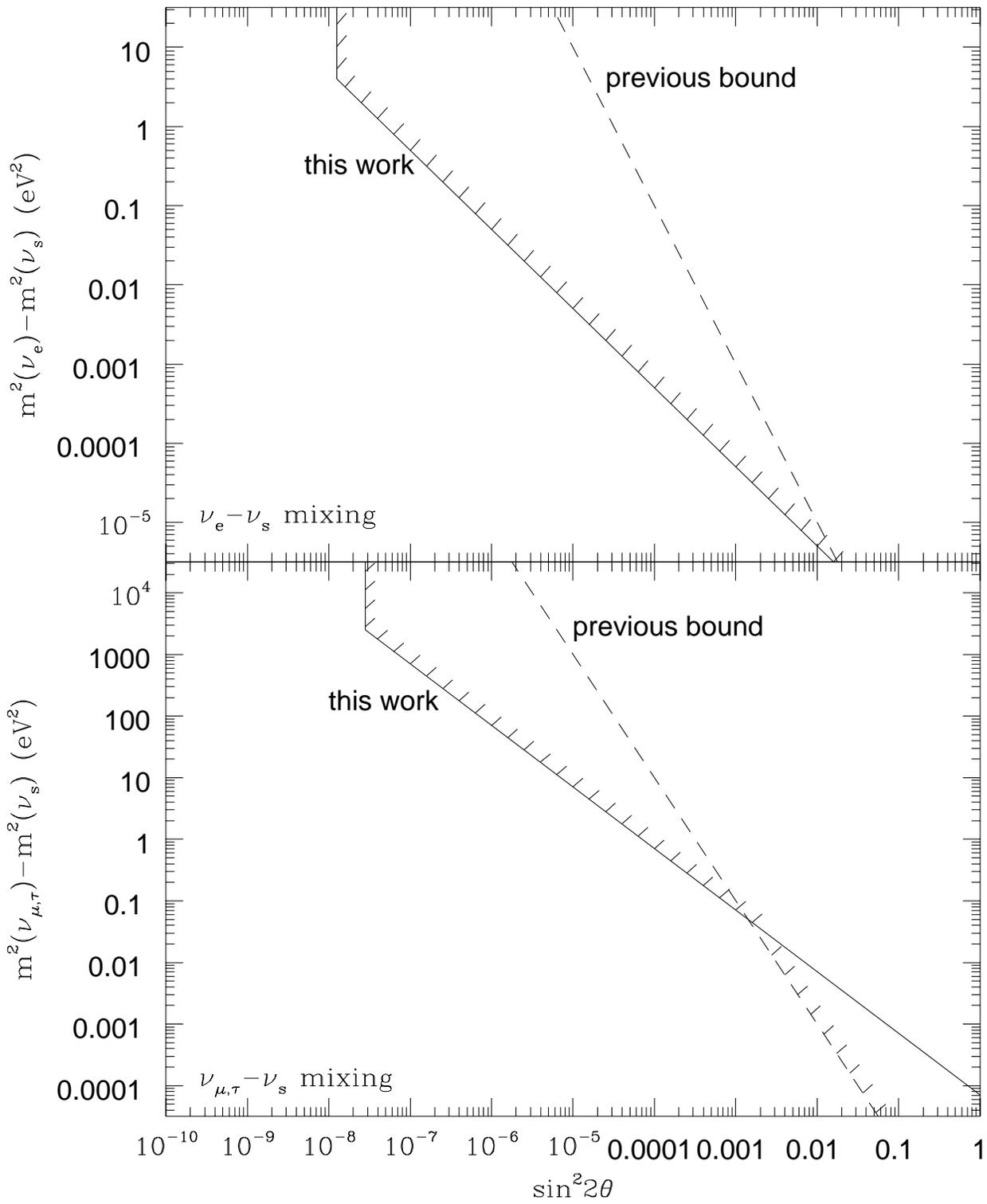
\end{document}